\documentclass[prx, notitlepage, twocolumn,nofootinbib,superscriptaddress
]{revtex4-1}
\usepackage[colorlinks=true,citecolor=blue,linkcolor=blue]{hyperref}
\usepackage{graphicx}
\usepackage{color}
\usepackage{dcolumn}
\usepackage{bm}
\usepackage{mathrsfs}
\usepackage{amsmath}
\usepackage{amssymb}
\usepackage{amsthm}
\usepackage{amsfonts}
\usepackage{enumerate}
\usepackage{latexsym}
\usepackage{hyperref}
\usepackage{centernot}
\usepackage{multirow}

\begin{document}

\title{Topological Materials Discovery By Large-order Symmetry
Indicators}

\begin{abstract}
Crystalline symmetries play an important role in the classification of band
structures, and the rich variety of spatial symmetries in solids leads to
various topological crystalline phases (TCPs). However, compared with
topological insulators and Dirac/Weyl semimetals, relatively few realistic
materials candidates have been proposed for TCPs. Based on our recently
developed method for the efficient discovery of topological materials using
symmetry indicators, we explore topological materials in five space groups
(i.e. $\mathcal{SG}s\mathbf{87},\mathbf{140},\mathbf{221,191},\mathbf{194}$%
), which are indexed by large order strong symmetry based indicators ($%
\mathbb{Z}_{8}$ and $\mathbb{Z}_{12}$) allowing for the realization of
several kinds of gapless boundary states in a single compound. We predict
many TCPs, and the representative materials include: Pt$_{3}$Ge($\mathcal{SG}%
\mathbf{140}$), graphite($\mathcal{SG}\mathbf{194}$), XPt$_{3}$ ($\mathcal{SG%
}\mathbf{221}$,X=Sn,Pb), Au$_{4}$Ti ($\mathcal{SG}\mathbf{87}$) and Ti$_{2}$%
Sn ($\mathcal{SG}\mathbf{194}$). As by-products, we also find that AgXF$_{3}$
($\mathcal{SG}\mathbf{140}$,X=Rb,Cs) and AgAsX ($\mathcal{SG}\mathbf{194}$%
,X=Sr,Ba) are good Dirac semimetals with clean Fermi surface. The proposed
materials provide a good platform to study the novel properties emerging
from the interplay between different types of boundary states.
\end{abstract}

\date{\today }
\author{Feng Tang}
\affiliation{National Laboratory of Solid State Microstructures and School
of Physics, Nanjing University, Nanjing 210093, China} %
\affiliation{Collaborative Innovation Center of Advanced Microstructures,
Nanjing 210093, China}

\author{Hoi Chun Po }
\affiliation{Department of Physics, Harvard University, Cambridge, MA 02138,
USA}

\author{Ashvin Vishwanath}
\affiliation{Department of Physics, Harvard University, Cambridge, MA 02138,
USA}

\author{Xiangang Wan}
\affiliation{National Laboratory of Solid State Microstructures and School
of Physics, Nanjing University, Nanjing 210093, China} %
\affiliation{Collaborative Innovation Center of Advanced Microstructures,
Nanjing 210093, China}
\maketitle

\affiliation{National Laboratory of Solid State Microstructures and School
of Physics, Nanjing University, Nanjing 210093, China}
\affiliation{Collaborative Innovation Center of Advanced Microstructures,
Nanjing 210093, China}

\affiliation{Department of Physics, Harvard University, Cambridge, MA 02138,
USA}

\affiliation{Department of Physics, Harvard University, Cambridge, MA 02138,
USA}

\affiliation{National Laboratory of Solid State Microstructures and School
of Physics, Nanjing University, Nanjing 210093, China}
\affiliation{Collaborative Innovation Center of Advanced Microstructures,
Nanjing 210093, China}

%
%
%
%
%
%
%

Since the discovery of two-dimensional (2D) and three-dimensional (3D)
topological insulators (TIs), band topology in condensed-matter materials
has attracted broad interest owing to their rich scientific implications and
potential for technological applications \cite{review-1,review-2}. Described
by $\mathbb{Z}_{2}$ topological invariant(s), time-reversal ($\mathcal{T}$)
invariant TIs are characterized by an insulating gap in the bulk and $%
\mathcal{T}$-protected gapless modes on the boundary of the system \cite%
{review-1,review-2}. Soon after the discovery of TIs, it was realized that
symmetry plays an important role in the classifications of topological
phases. Based on the absence or presence of $\mathcal{T}$, particle-hole or
chiral symmetry, insulators and superconductors have been classified into
the so-called ten-fold periodic table \cite{rmp}.

In addition to the aforementioned internal symmetries, the topological
classification of band structures has also been extended to include
crystalline symmetries \cite{TCI-1,TCI-2,TCI-3}, and due to the vast array
of crystal symmetries (encapsulated by the 230 crystalline space group ($%
\mathcal{SG}$)), massive topological crystalline phases (TCPs) have been
proposed, such as: mirror Chern insulator \cite{MCI}, quantized electric
multipole insulators \cite{electric multipole insulators}, high order
topological insulator \cite{electric multipole
insulators,High-order-1,High-order-2,High-order-3,High-order-4}, hourglass
fermions \cite{hourglass}, nodal-chain metals \cite{nodal-chain},
unconventional quasiparticles with three-fold (or higher) band degeneracies
\cite{New-ferimon}\ etc. Very recently, by exploiting the mismatch between
the real and momentum-space descriptions of the band structure, novel forms
of band topology in the 17 wallpaper groups \cite{wallpaper-Kane}, the 230 $(%
\mathcal{SG}s)$ for nonmagnetic compounds \cite{Po,QCT}, and the 1651
magnetic $\mathcal{SG}s$ for magnetic materials have been proposed \cite%
{Po-2}. Currently, the standard way for finding topological materials is
based on the evaluation of various topological invariants \cite%
{TCI-1,TCI-2,TCI-3,MCI,hourglass,wallpaper,High-order-1,Fu-Kane,FangFu,Fang-1,Fang-2,Fang-3,Fang-4,TCI-4,TCI-6,Khalaf-1,Max,Khalaf}%
. As the calculation for topological invariants is usually a time-consuming
task, the finding of any realistic topological materials is typically taken
as a big success \cite%
{review-1,review-2,TCI-3,MCI,hourglass,nodal-chain,New-ferimon,Weyl-1,Weyl-2,Weyl-3,Ca2As}%
. The discovered topological compounds represents a very small fraction of
the experimentally synthesized materials tabulated in structure databases
\cite{ICSD}. Thus, the search for new TCPs with novel properties is of both
fundamental and technological importance, and we address this issue by our
recently developed method for diagnosing topological materials \cite%
{TCP-Paper-1}.

Our method integrates the recently established theory of symmetry indicators
(SI) of band topology into first-principle band-structure calculations \cite%
{wallpaper-Kane,Po,QCT,Po-2,TCP-Paper-1}. As shown in Ref.\cite{TCP-Paper-1}%
, after standard electronic structure calculation, one needs only to
calculate the representations of filled energy bands at high-symmetry
points, i.e. $n_{\mathbf{k}}^{\alpha }$ which can be written as a formal
vector: $\mathrm{\mathbf{n}}=(\nu ,n_{\mathbf{k_{1}}}^{1},n_{\mathbf{k_{1}}%
}^{2},\ldots )$, where $\nu $ is the total number of the filled energy
bands, the subscript $\mathbf{k}_{1},\mathbf{k_{2}},\ldots ,\mathbf{k}_{N}$
denotes the high symmetry point (HSP) in the BZ, the superscript $1,2,\ldots
,\alpha _{i},\ldots $ refers to the irreducible representation (irrep) of
little group at $\mathbf{k}_{i}$ point ($\mathcal{G}_{\mathbf{k}_{i}}$), and
$n_{\mathbf{k}_{i}}^{\alpha _{i}}$ means the number of times an $\alpha _{i}$%
\ irrep of $\mathcal{G}_{\mathbf{k}_{i}}$ appears among the filled bands.

It was realized that the set of vectors $\mathrm{\mathbf{n}}$\ forms an
abelian group \cite{wallpaper-Kane, Po}. Moreover, for every $\mathcal{SG}$,
there exists $d_{\mathrm{AI}}$ atomic insulator (AI) basis vectors ($\mathbf{%
a}_{i},i=1,2,\ldots ,d_{\mathrm{AI}}$) containing information of the group
structure for the symmetry-based indicator (SI), denoted by $X_{\mathrm{BS}}$
in Ref.\ \onlinecite{Po}, according to the possible common factor $C_{i}$
for $\mathbf{a}_{i}$ \cite{Po}. One can always expand any vector $\mathrm{%
\mathbf{n}}$\ with respect to the AI basis vectors $\mathbf{a}_{i}$: $%
\mathrm{\mathbf{n=}}\overset{d_{AI}}{\underset{i=1}{\sum }}q_{i}\mathbf{a}%
_{i} $. The expansion coefficients of $\mathrm{\mathbf{n}}$ on the AI basis
can be classified into three cases \cite{TCP-Paper-1}: \textbf{Case 1}: the
expansion coefficients ${q_{i}}^{\prime }s$ are all integers; such materials
might be adiabatically connected to a trivial atomic insulator, and so we do
not consider materials in this case. \textbf{Case 2}: the expansion
coefficients ${q_{i}}^{\prime }s$ are not all integers, but all ${q_{i}C_{i}}%
^{\prime }s$ are integers; such materials are necessarily topological and
the results of $(q_{i}C_{i}\quad \mathrm{mod}\quad C_{i})$, gives the
nonvanishing SI \cite{TCP-Paper-1}. \textbf{Case 3}: the ${q_{i}C_{i}}%
^{\prime }s$ are not all integers; such systems are (semi-)metallic.
Specifically, if $n_{\mathbf{k}_{i}}^{\alpha }$ is non-integer then there is
band crossing happens at $k_{i}$ point; on the other hand, if all the ${n_{%
\mathbf{k}}^{\alpha }}^{\prime }s$ are integers, then there must be band
crossing in high symmetry line or plane \cite{TCP-Paper-1}.

%

There are various topological invariants, which correspond to different
kinds of band topology. During the regular searches for topological
materials, one need to decide which topological invariant to evaluate.
Moreover, usually the calculations for the topological invariants are a
computationally heavy task. In stark contrast to conventional
target-oriented searches, our algorithm does not presuppose any specific
phase of matter. Based on the expansion coefficients, which are very easy to
calculate, one can quickly identify the topological (semi-)metals,
topological insulators and topological crystalline insulators \cite%
{TCP-Paper-1}. The high efficiency of our method has been demonstrated in
Ref. \onlinecite{TCP-Paper-1}, in which we discuss many topological
materials discovered based on their nontrivial index in space groups with $%
\mathbb{Z}_{2}$ or $\mathbb{Z}_{4}$ strong factor in the SI group. %

One of the hallmarks of topological phases is the bulk-boundary
correspondence \cite{review-1,review-2}, and different types of topological
boundary states, such as Dirac surface states, hourglass surface states, and
more recently hinge states, have been proposed. Thus, finding realistic
materials with the coexistence of various topological boundary states is a
very interesting issue. In this work we focus on $\mathcal{SG}s$ with the
larger strong factor in the SI group, $X_{\mathrm{BS}}^{s}$, i.e., $\mathbb{Z%
}_{8}$ and $\mathbb{Z}_{12}$, where various types of band topology are
expected \cite{Po,Khalaf,Fang-3}. These SIs are realized in $\mathcal{SG}s$
with a high-degree of coexisting symmetries, such as (roto-)inversion,
mirror reflection, screw, and glide etc. There are in total 12 and 6 $%
\mathcal{SG}s$ with strong $\mathbb{Z}_{8}$ and $\mathbb{Z}_{12}$ SI factor
group, respectively \cite{Po}. Focusing on five $\mathcal{SG}s$ with $%
\mathbb{Z}_{8} $ or $\mathbb{Z}_{12}$ strong SI group (i.e. $\mathcal{SG}s~%
\mathbf{87},\mathbf{140},\mathbf{221,191},\mathbf{194}$), we search for
interesting TCPs in a single sweep of the crystal database \cite{ICSD} using
the method delineated in Ref.\ \onlinecite{TCP-Paper-1}. We only consider
spin-orbital coupled non-magnetic materials with $\leq 30$ atoms in their
primitive unit cell. We find a large number of TCPs with reasonably clean
Fermi surfaces. In the following, we present and discuss six representative
topological crystalline insulators (TCIs), and list others in Tables \ref%
{tab:z8} and \ref{tab:z12}. The 4 good Dirac semimetals are discussed in the
end.

\begin{table}[tbp]
\centering%
\begin{tabular}{|c|c|c|}
\hline
$\mathcal{SG}$ & $X_{\mathrm{BS}}$ & material(SI) \\ \hline
$\mathbf{87}$ & $\mathbb{Z}_{2}\times \mathbb{Z}_{8}$ & ~~~~{%
\color{red}{Au$_4$Ti}}~~~~ \\ \hline
$\mathbf{140}$ & $\mathbb{Z}_{2}\times \mathbb{Z}_{8}$ & ~~~~{%
\color{red}{Pt$_3$Ge}}(04),SiTa$_{2}$(11)~~~~ \\ \hline
$\mathbf{221}$ & $\mathbb{Z}_{4}\times \mathbb{Z}_{8}$ &
\begin{tabular}{m{5cm}}
AlX(X=Sc,Y)(03) \\
XB$_{6}$(X=Ca,Sr,Ba)(03) \\
BeTi(03), CaPd(20),CsPbBr$_{3}$(23) \\
CsGeBr$_{3}$(23),CsSnI$_{3}$(23) \\
Ca$_{3}$PbO(22),{\color{red}XPt$_{3}$(X=Pb,Sn)}(34)%
\end{tabular}
\\ \hline
\end{tabular}%
\caption{The topological crystalline (TC) insulating materials for $\mathcal{%
SG}s\mathbf{87},\mathbf{140}$ and $\mathbf{221}$. These $\mathcal{SG}s$ all
own the same strong SI factor group: $\mathbb{Z}_{8}$ but with different
other weak SI factor groups. The red color denotes the materials carefully
discussed in this work.}
\label{tab:z8}
\end{table}

\begin{table}[tbp]
\centering
\begin{tabular}{|c|c|c|}
\hline
$\mathcal{SG}$ & $X_{\mathrm{BS}}$ & material(SI) \\ \hline
$\mathbf{191}$ & $\mathbb{Z}_6\times\mathbb{Z}_{12}$ & XB$_2$%
(X=Mg,Ca)(52),SrB$_2$(15),Ti(33) \\ \hline
$\mathbf{194}$ & $\mathbb{Z}_{12}$ &
\begin{tabular}{m{5cm}}
AlLi(4),AlC$_2$Ta$_3$(1),Ca$_2$NI(3) \\
{\color{red}{graphite}(4)},Na$_2$CdSn(4) \\
MgPo(1),SiSr$_2$(1) \\
{\color{red}{Ti$_2$Sn}}(6)%
\end{tabular}
\\ \hline
\end{tabular}%
\caption{The TC insulating materials for $\mathcal{SG}s\mathbf{191}$ and $%
\mathbf{194}$. These $\mathcal{SG}s$ all own the same strong SI factor
group: $\mathbb{Z}_{12}$ but with different other weak SI factor groups. The
red color denotes the materials carefully discussed in this work.}
\label{tab:z12}
\end{table}

\section{$\mathbb{Z}_8$ nontrivial TCI: P\lowercase{t}$_3$G\lowercase{e}}

\begin{figure}[tbp]
\includegraphics[width=7.5cm]{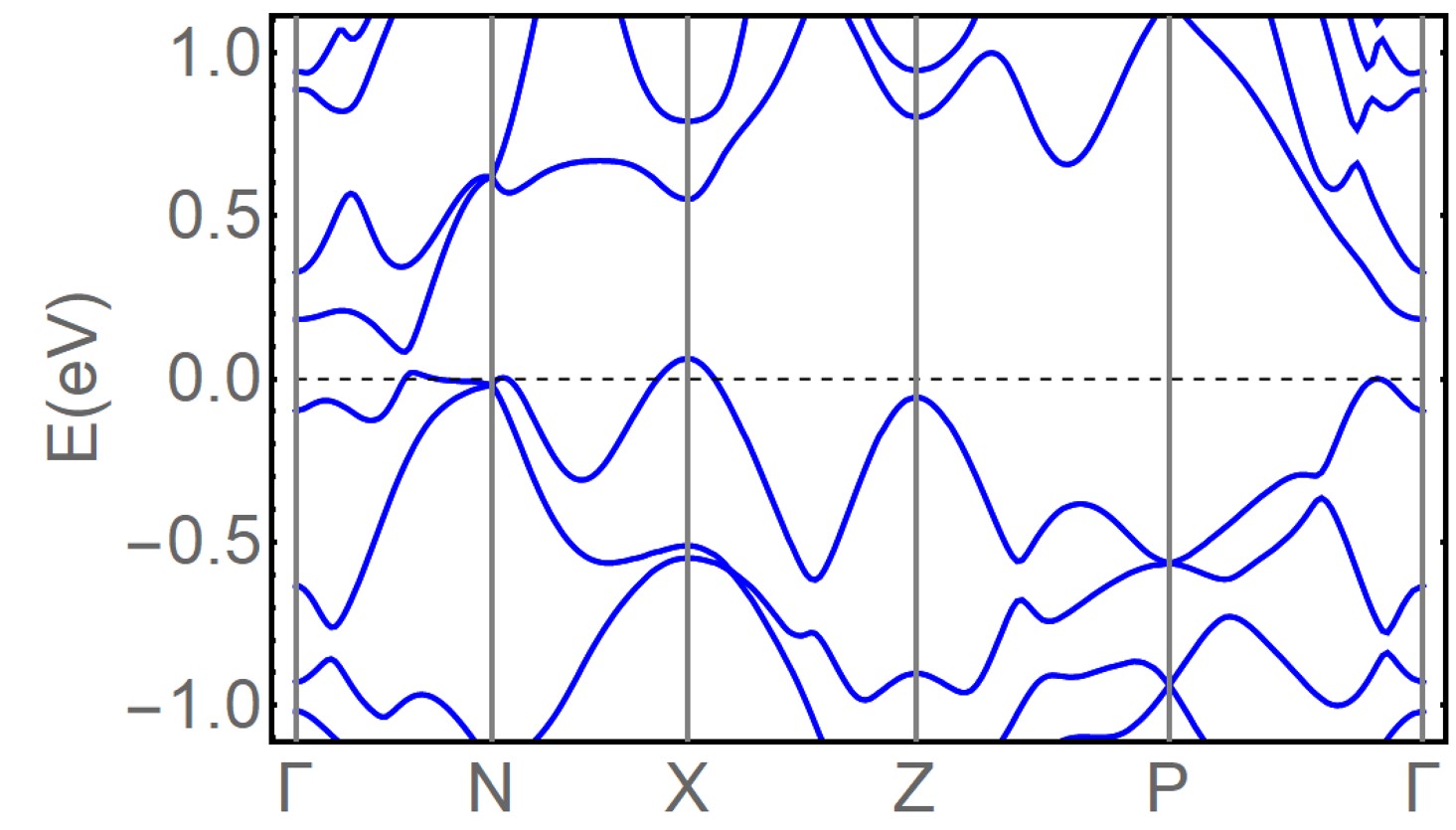}\newline
\caption{The electronic band plot of TCI Pt$_{3}$Ge within $\mathcal{SG}%
\mathbf{140}$.}
\label{fig:Pt3Ge}
\end{figure}

We first search the nonsymmorphic $\mathcal{SG}\mathbf{140}$ ($I4/mcm$),
which has 7 AI basis vectors: $\mathbf{a}_{i}^{\mathcal{SG}\mathbf{140}%
},i=1,2,\ldots ,7$. Only 2 AI basis vectors (we label them as $\mathbf{a}%
_{6}^{\mathcal{SG}\mathbf{140}}$ and $\mathbf{a}_{7}^{\mathcal{SG}\mathbf{140%
}}$) have a common factor: 2 and 8 respectively. Thus, the SI group of $%
\mathcal{SG}\mathbf{140}$ is $X_{\mathrm{BS}}^{\mathcal{SG}\mathbf{140}}=%
\mathbb{Z}_{2}\times \mathbb{Z}_{8}$. There are 158 materials with small
unit cell and without magnetic atoms. The calculations based on our method
\cite{TCP-Paper-1} identified 77 topological nontrivial compounds, with 27
belonging to case 2 while 50 being topological (semi-)metals indicated by
case 3. Further filtering by Fermi level criteria which requires that the
energy bands around the Fermi level are clean as far as possible, we list
the relatively good materials in Table \ref{tab:z8}. In the following, we
take Pt$_{3}$Ge \cite{Pt3Ge} as the example to analyze the detailed
topological properties.

Pt$_{3}$Ge crystallize in the body-centered tetragonal structure \cite{Pt3Ge}%
, where Ge occupies the $4b$ Wyckoff position, and Pt's occupying two sets
of inequivalent sites in the $4a$ and $8h$ Wyckoff positions. There are in
total 68 valence electrons in the primitive unit cell. Based on \textit{ab
initio} calculation, we calculate the irrep multiplicities ${n_{\mathbf{k}%
}^{\alpha }}^{\prime }s$ for all the high symmetry points and all the
corresponding irreps $\alpha $ for the 68 valence bands. We then expand this
calculated\ vector $\mathrm{\mathbf{n}}$ on the 7 AI basis vectors: $\mathrm{%
\mathbf{n}}=\sum_{i=1}^{7}q_{i}\mathbf{a}_{i}^{\mathcal{SG}\mathbf{140}}$,
and obtain $q=(8,0,0,1,2,1,-\frac{1}{2})$. Thus this material belongs to
case 2, and is a TCI with SI being (0,4). As seen from the electronic band
plot in Fig. \ref{fig:Pt3Ge}, this material has large direct gaps through
the $k$ path.

While from SI alone we can ascertain that Pt$_{3}$Ge is a TCI, to resolve
the concrete form of band topology it displays we have to evaluate
additional topological indices. First, we note that from the SI we can infer
all the Fu-Kane parity criterion \cite{Fu-Kane} is silenced, i.e., the
material cannot be a strong nor weak TI.
The enriched inversion invariant $\delta _{i}$ \cite{Fang-3,Khalaf} ($\delta
_{i}\equiv (\kappa _{1}\quad \mathrm{mod}\quad 4)/2$) is also vanishing.
Thus this material has boundary states protected by symmetry operation
containing $n-$fold axis ($n>1$) \cite{Fang-3,Khalaf}. Due to the rich point
symmetry operations in $\mathcal{SG}\mathbf{140}$ (whose point group is $%
D_{4h}$), several topological phases may occur \cite{Fang-3,Khalaf}. We thus
evaluate the mirror Chern numbers for the (001) (Miller indices with respect
to the conventional lattice basis vectors) and (110)-mirror planes by first
principles calculations. Our numerical results show that they are also all
vanishing. The glide, screw and $S_{4}$ invariant is thus nonvanishing \cite%
{Fang-3, Khalaf}. Thus it would have glide protected hourglass surface
states in (100) glide symmetric planes as the corresponding invariant is 1.
The (001) screw invariant is 1 thus it would protect gapless hinge states
along the $\mathbf{c}$ direction.

\section{$\mathbb{Z}_{12}$ nontrivial TCI: graphite}

We also searched the 492 materials with $\mathcal{SG}\mathbf{194}$($%
P6_{3}/mmc$, whose point group is $D_{6h}$) in the database\cite{ICSD}.
There are 52 and 254 materials belonging to cases 2 and 3 respectively. It
is worth emphasizing that our results indicate that graphite \cite{C} is
potentially a nontrivial insulator.

It is well-known that graphene (i.e. monolayer of graphite) exhibits 2D
massless Dirac excitation near $K/K^{\prime }$ points \cite{Graphene-RMP}.
The spin-orbit (SO) coupling (although small), opens a topological gap ($%
\sim $0.0008 meV \cite{Graphene-gap}), making it, in principle, a 2D
topological insulator \cite{Kane-2005}. The direct stacking of graphene will
then lead to a weak topological insulator.
The discovery of crystalline-symmetry-protected band topology in graphite,
the ABABABAB$\ldots $ Bernal stacking of graphene, demonstrates the
possibilities of discovering topological materials even among the simplest
elemental materials.
We thus present a detailed discussion in the following.

\begin{figure}[tbp]
\includegraphics[width=7.5cm]{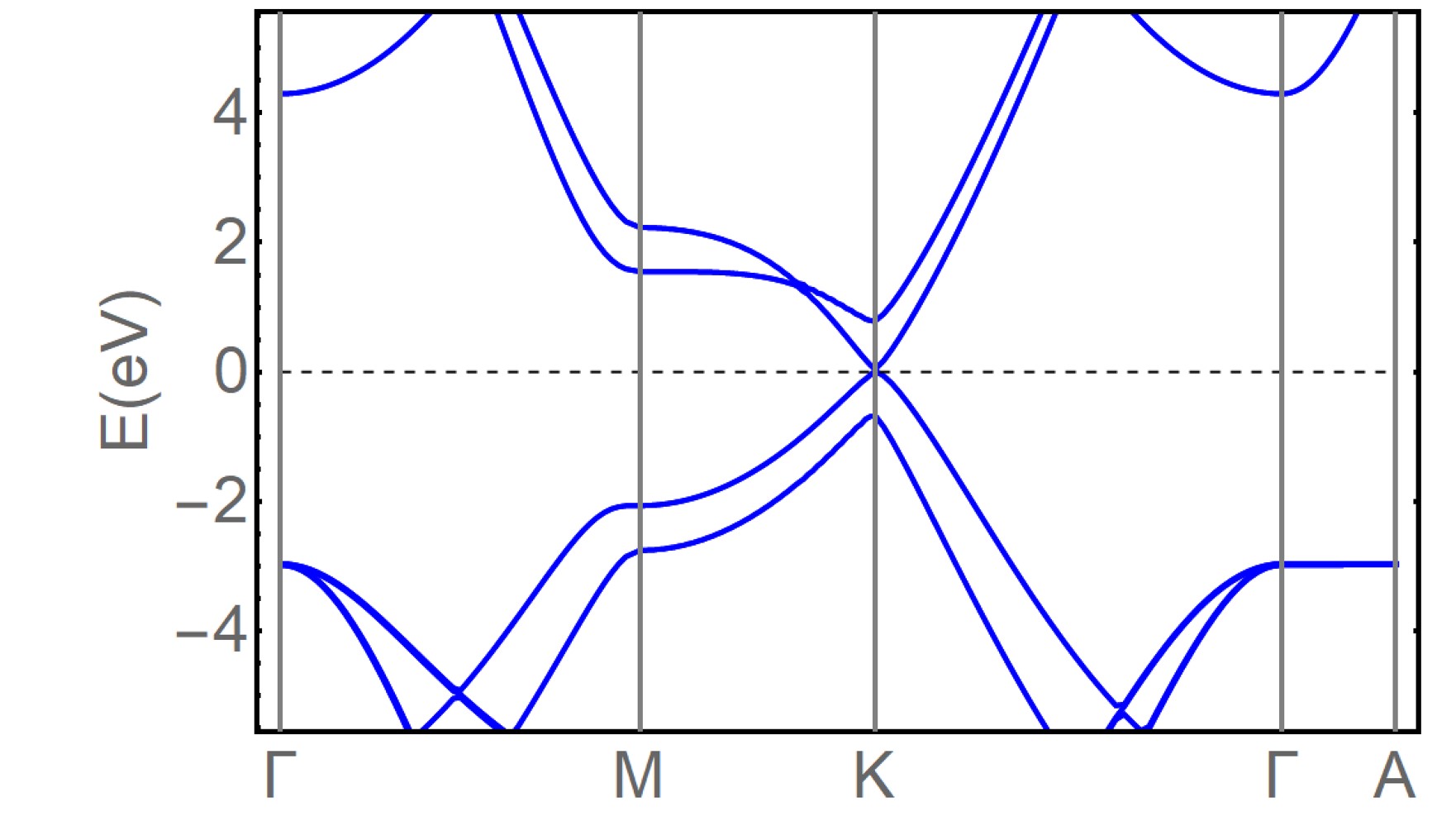}\newline
\caption{The electronic band plot of TCI graphite within $\mathcal{SG}%
\mathbf{194}$.}
\label{fig:C}
\end{figure}

The $\mathcal{SG}\mathbf{194}$ owns 13 AI basis vectors $\mathbf{a}_{i}^{%
\mathcal{SG}\mathbf{194}},i=1,2,\ldots,13$, where only the last one has a
common factor, which is 12. Thus $X_{\mathrm{BS}}^{\mathcal{SG}\mathbf{194}}=%
\mathbb{Z}_{12}$. The 16 valence bands in graphite are found to have the
expansion coefficients $q=(2,0,-1,-1,-1,-1,1,3,\frac{1}{3})$ on the AI
basis. Thus the SI for graphite is $4\in \mathbb{Z}_{12}$. The band
structure is shown in Fig. \ref{fig:C}, where SO coupling opens a small gap
(around 0.025 meV) at the $K$ point. The inversion invariant $\delta _{i}$
and three Fu-Kane weak topological invariants \cite{Fu-Kane} are found to be
all vanishing. We then calculate the (001)-mirror Chern number and find that
it is $-2$. Thus there would be gapless Dirac surface states in the (001)
mirror symmetric planes. In order to ascertain graphite's nontrivial
topology, we then calculate the ($\bar{1}20$) plane's mirror Chern, and find
that it is vanishing. Then it would has 6-fold screw protected hinge states
\cite{Fang-3,Khalaf}. It may also have glide and rotation protected surface
states as dictated by the nonvanishing $\{c_2^{1\bar{1}0}|000\}$ (in Seitz
notation where the superscript of the point operation part denotes that
rotation axis and the subscript denotes the rotation angle) rotation
invariant and (010) glide invariant \cite{Khalaf, Fang-3}. While graphite is
generally associated with small Fermi pockets, Ref. \cite{graphite-gap}
proposed, based on the observation of a semiconducting gap in small samples
of Bernal graphite, that these may arise from extrinsic effects. Thus ,
further experimental work would be of great interest.

\section{The Other discovered TCIs}

\begin{figure*}[tbp]
\includegraphics[width=16cm]{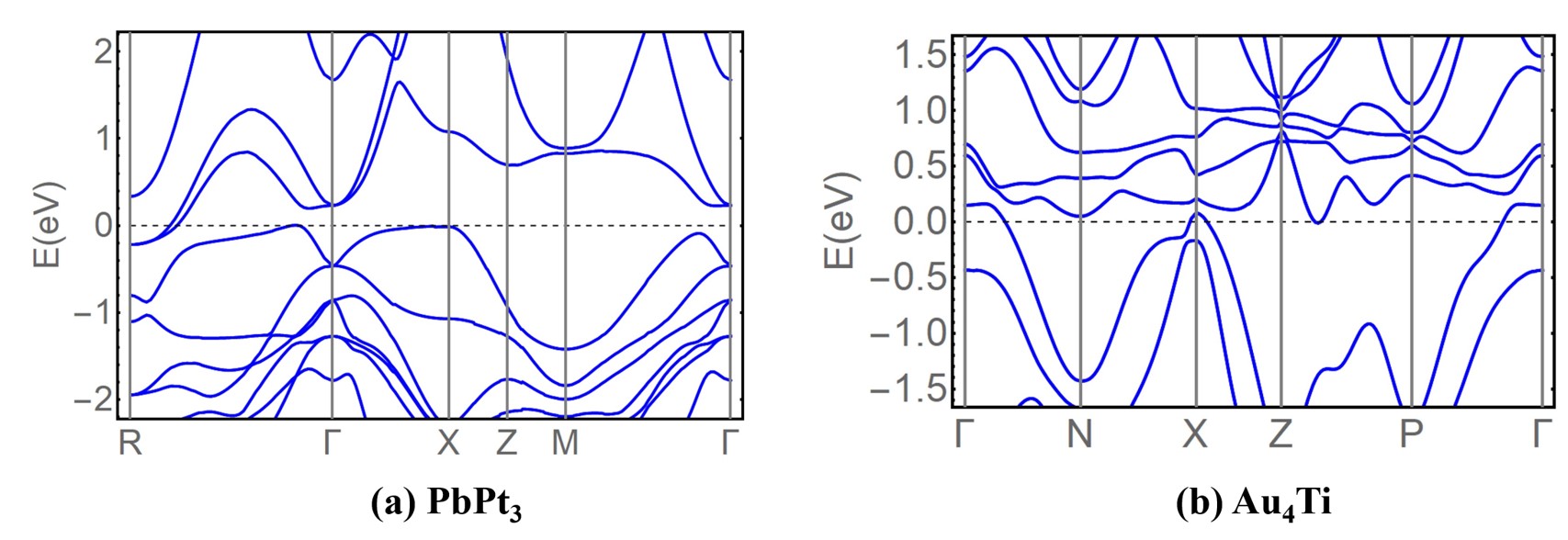}\newline
\caption{The electronic band plots of TCIs PbPt$_3$ within $\mathcal{SG}%
\mathbf{221}$ and Au$_4$Ti within $\mathcal{SG}\mathbf{87}$ .}
\label{fig:PbPt3+Au4Ti}
\end{figure*}

\subsection{Weak TI coexisting with TCI in PbPt$_{3}$($\mathcal{SG}\mathbf{%
221}$) and Au$_{4}$Ti($\mathcal{SG}\mathbf{87}$)}

The above two TC materials both have vanishing inversion and weak
topological invariants. We also discover two materials, i.e. PbPt$_{3}$ in $%
\mathcal{SG}\mathbf{221}$ and Au$_{4}$Ti in $\mathcal{SG}\mathbf{87}$ which
have three weak topological indices \cite{Fu-Kane} $\nu _{i}=1$ for $i=1,2,3$%
: they have inversion topological invariant $\delta _{i}$ \cite%
{Fang-3,Khalaf} equal to 0 and 1, respectively.

PbPt$_{3}$ crystallizes in the cubic structure with a primitive Bravais
lattice. The electronic band structure is shown in Fig. \ref{fig:PbPt3+Au4Ti}%
. The material has 34 valence electrons in the unit cell. The calculated ${%
n_{\mathbf{k}}^{\alpha }}^{\prime }s$ for these 34 bands can be expanded on
the 14 AI basis vectors of $\mathcal{SG}\mathbf{221}$, and the expansion
coefficients are $q=(0,0,0,0,0,-1,1,1,0,1,-1,-1,-\frac{1}{4},-\frac{1}{2})$.
The last two AI basis vectors own a common factor 4 and 8, respectively.
Thus the SI is $(3,4)\in \mathbb{Z}_{4} \times \mathbb{Z}_{8}$. The parity
calculations show that it is a weak topological insulator \cite{Fu-Kane}. We
also calculate the two mirror Chern numbers for (001) mirror plane ($k_{z}=0$
or $\pi $), and find that they are both equal to $-1$. At the same time, the
screw invariant of $\{c_{2}^{011}|0\frac{1}{2}\frac{1}{2}\}$ is 1. Thus this
material can host protected hinge and surface states at the same time.

Au$_{4}$Ti \cite{Au4Ti} crystallizes in $\mathcal{SG}\mathbf{87}$ ($I4/m$),
where Au and Ti occupy $8h$ and $2a$ Wyckoff positions respectively. This
material is found to belong to case 2. We calculate the parities and find
that its strong topological invariant \cite{Fu-Kane} and inversion invariant
\cite{Fang-3,Khalaf} are both vanishing i.e., $\nu _{0}=\delta_{i}=0$ while $%
\nu _{1}=\nu _{2}=\nu _{3}=1$, so it is a weak TI. Besides, the newly
introduced invariant $\Delta$ \cite{Khalaf} is found to be 4 (mod 8). Our
first principles calculations also show that the mirror Chern number for the
(001)-plane is vanishing. Thus it would allow glide protected hourglass
surface states in glide $\{m_{001}|\frac{1}{2}\frac{1}{2}0\}$ symmetric
plane. It can also host hinge states along (001) direction which are
protected by screw $\{c_{2}^{001}|00\frac{1}{2}\}$ or $\{c_{4}^{001}|00\frac{%
1}{2}\}$.

\subsection{TCI Ti$_2$Sn in $\mathcal{SG}\mathbf{194}$}

\begin{figure}[tbp]
\includegraphics[width=7.5cm]{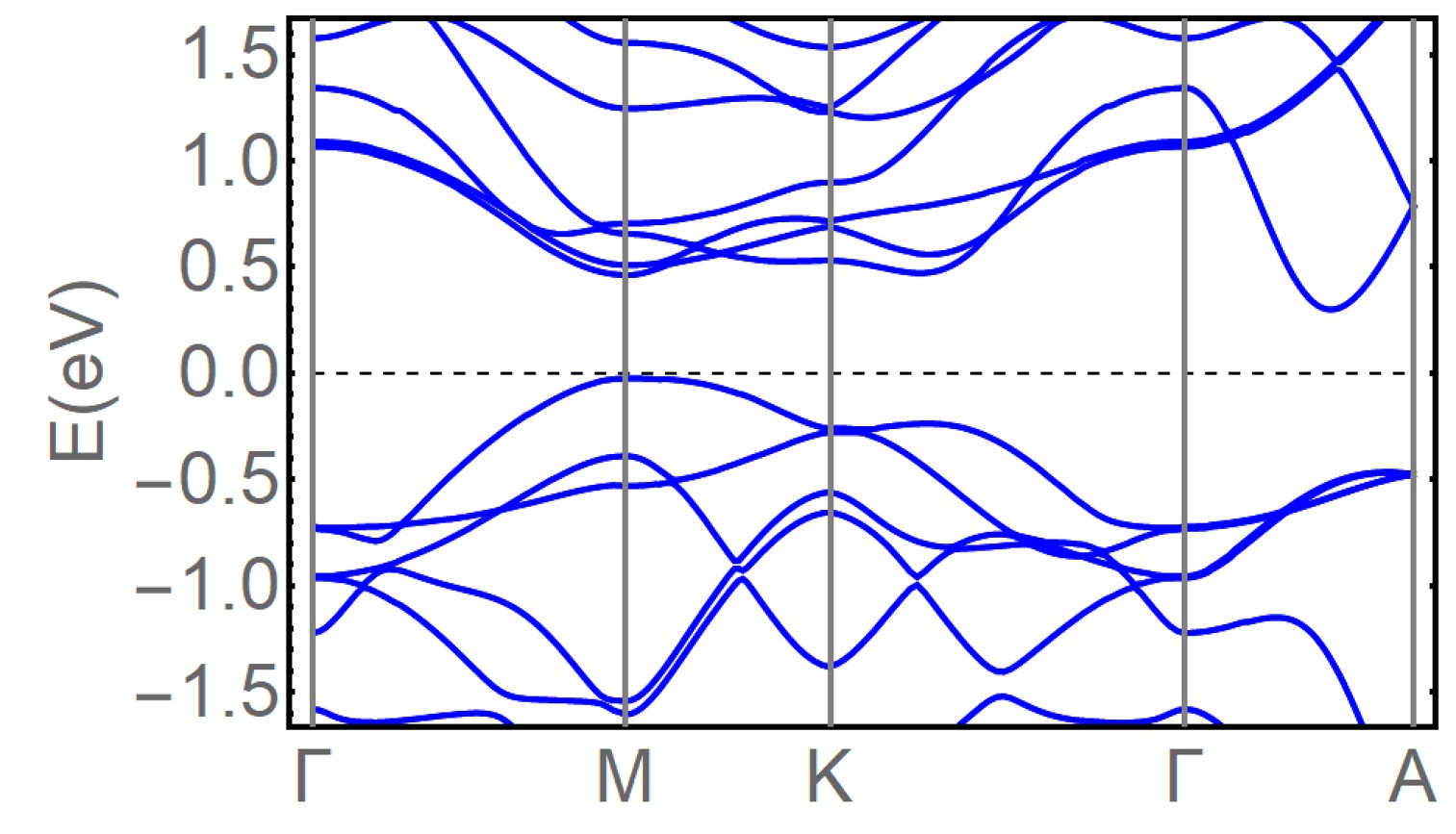}\newline
\caption{The electronic band plot of TCI Ti$_{2}$Sn within $\mathcal{SG}%
\mathbf{194}$.}
\label{fig:Ti2Sn}
\end{figure}

Ti$_{2}$Sn \cite{Ti2Sn} within $\mathcal{SG}\mathbf{194}$ is found to be a
TCI. It has large direct gaps everywhere except in a small area where there
are little electron and hole pockets. Our calculation show that the SI is
(6). Parity calculations show that the inversion invariant $\delta _{i}$
\cite{Fang-3,Khalaf} is 1 while the strong and weak topological invariants
\cite{Fu-Kane} $\nu _{0,1,2,3}$ are all vanishing. From first principles
calculation, we find the mirror Chern number for the ($\bar{1}$20) plane is
-4. This high mirror Chern number indicates that there should be multiple
Dirac cones in the ($\bar{1}20$) mirror symmetric plane. In order to
identify the band topology, we also calculate the mirror Chern number of the
(001) mirror plane, which is found to be vanishing. Thus it can accommodate
hourglass surface states in $\{m_{010}|00\frac{1}{2}\}$ or $\{m_{010}|\frac{1%
}{2}0\frac{1}{2}\}$ glide symmetric planes. $c_{2}$ around $(010)$ can also
protect surface Dirac cones. Besides, inversion and screw $\{c_{6}^{001}|00%
\frac{1}{2}\}$ can protect hinge states in corresponding hinges satisfying
the corresponding symmetries.

\section{topological semimetals}

\begin{figure*}[htb]
\includegraphics[width=16cm]{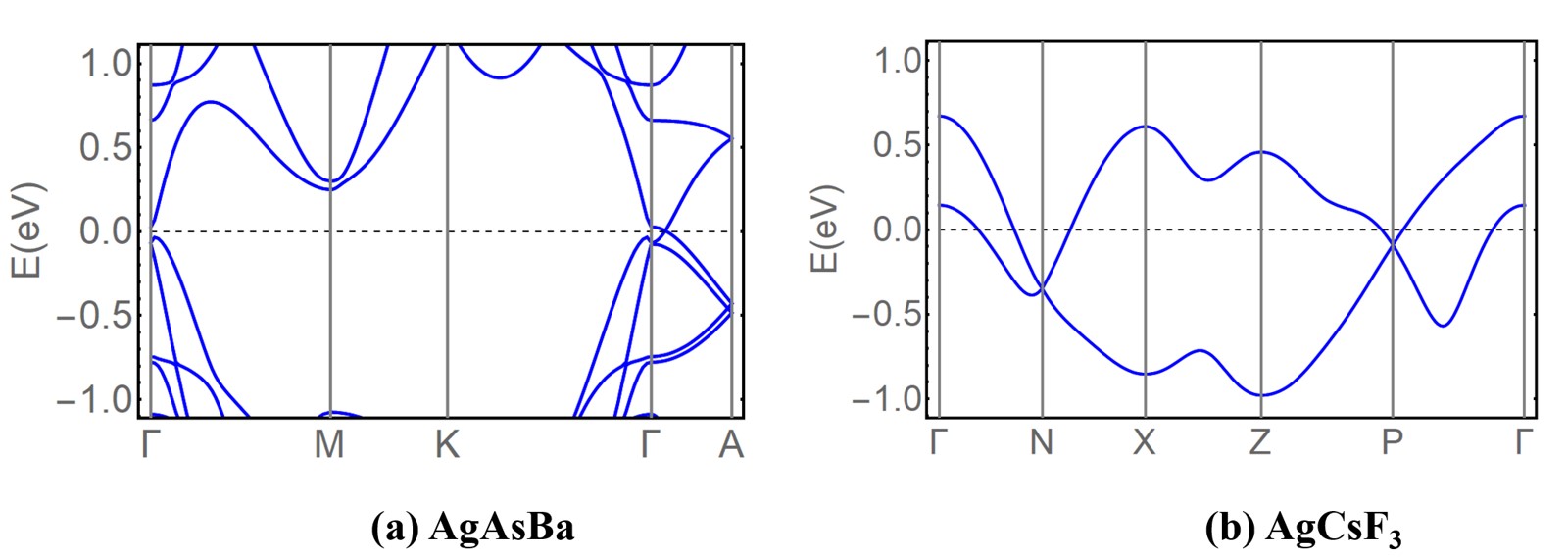}\newline
\caption{The electronic band plots of Dirac semimetals AgAsBa within $%
\mathcal{SG}\mathbf{194}$ and AgCsF$_{3}$ within $\mathcal{SG}\mathbf{140}$
. }
\label{fig:DSM}
\end{figure*}

\begin{figure*}[!htb]
\includegraphics[width=15cm]{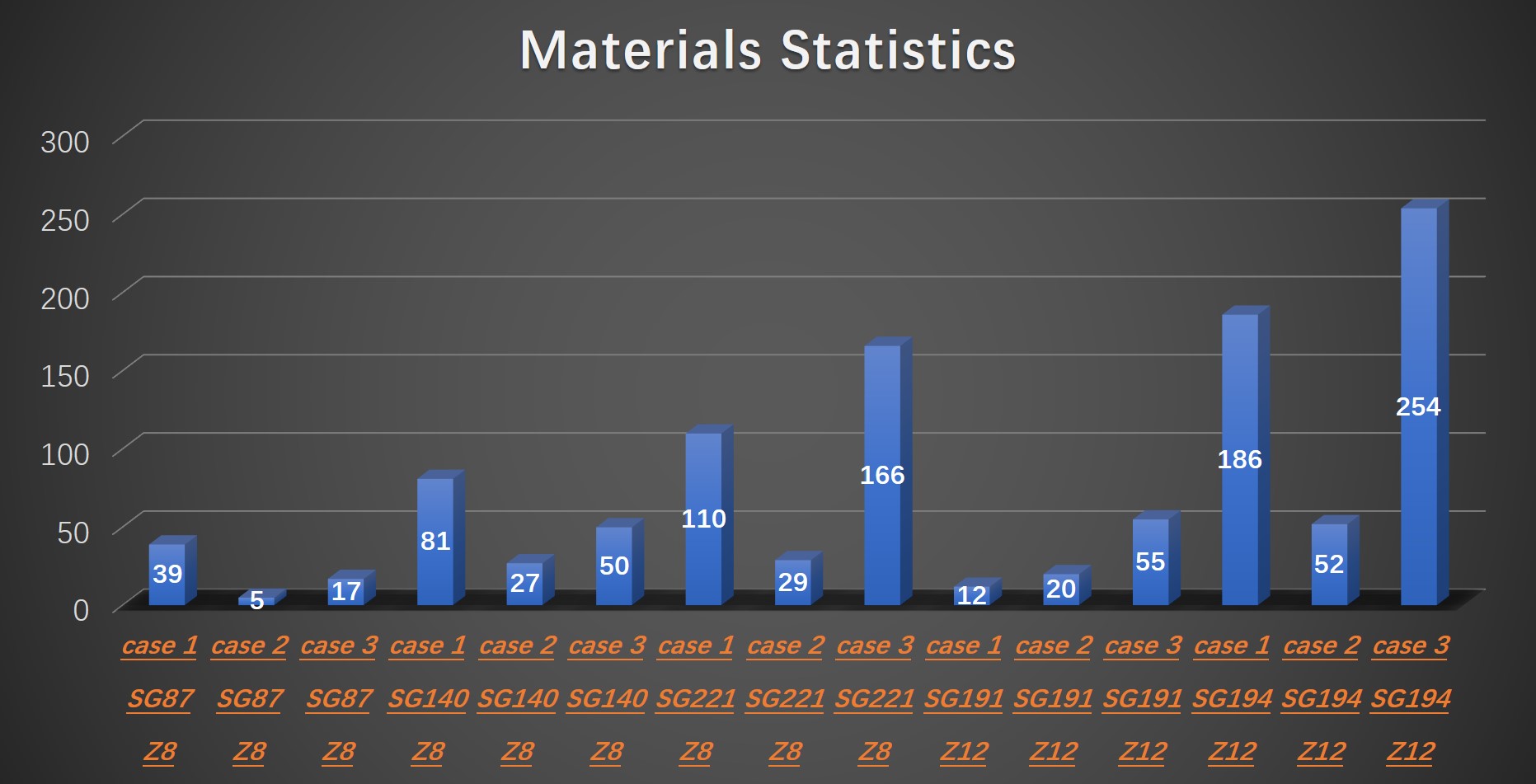}\newline
\caption{The statistics of topological materials search using the SI method.
The number in each bar indicates the number of materials for each case. Case
1 corresponds to the trivial situation, while case 2 and 3 correspond to
topological band structures and semimetals respectively. Requiring a minimal
Fermi surface further reduces the number of viable candidates. }
\label{fig:stat}
\end{figure*}
Other than the TCIs, our method can also filter out topological
(semi-)metals when the expansion coefficients belong to case 3. By further
requiring relatively clean Fermi surfaces, we identify AgXF$_3$\cite{AgXF3}%
(X=Rb,Cs,$\mathcal{SG}\mathbf{140}$) as good Dirac semimetals with Dirac
points pinned down to two high symmetry points ($P$ and $N$), and AgAsX\cite%
{AgAsSr,AgAsBa}(X=Sr,Ba,$\mathcal{SG}\mathbf{194}$) as Dirac semimetals with
symmetry-protected band crossing at high symmetry line, as shown in Fig. \ref%
{fig:DSM}. These two materials families realize the two sub-cases within
case 3 that we discussed. For the AgXF$_3$ family, the high symmetry points $%
P$ and $N$ both have only one 4-dimensional irrep while the filling cannot
be divided by 4. The filling-enforced Dirac points at $P$ or $N$ are
subjected to more symmetry restrictions than those for the Dirac points in
high symmetry line, and consequently the Dirac dispersion is more isotropic.
For the AgAsX family, in the high symmetry line $\Gamma$-$A$, the $\Delta_7$
and $\Delta_9$ band crossing each other, resulting a Dirac point protected
by $C_{6v}$. The Fermi level exactly threads the Dirac point.

\section{Conclusions and Perspectives}

In this work, based on our newly developed algorithm \cite{TCP-Paper-1}, we
search for topological materials indicated by $\mathbb{Z}_{8}$ and $\mathbb{Z%
}_{12}$ strong factors in the SI groups. Focusing on $\mathcal{SG}s\mathbf{%
87,140,221,191,194} $, we predict many materials, which exhibit coexistence
of various gapless boundary states due to the rich combination of various
symmetry operators in these highly symmetric $\mathcal{SG}s$. Breaking the
symmetry operation directly affects (move or even gap)\ the gapless
topological boundary state, thus one may easily tune the novel properties of
these predicted topological materials through strain or boundary decoration.

It is worth mentioning that the electronic topological phenomenon is
widespread in real materials and as shown in the Fig. \ref{fig:stat},
majority of the materials in the five $\mathcal{SG}s$ we scanned belong to
topological phases indicated by cases 2 and 3. In this manuscript, we only
discuss the materials with clean Fermi surfaces, since in these materials we
expect the transport properties to be dominated by the topologically
non-trivial states. Our scheme also finds some good metal with big Fermi
surfaces possessing non-zero SI. One good example is MgB$_{2}$. It is
interesting to contemplate on the possible interplay between its
superconductivity \cite{MgB2} and band topology.

We hope that our proposed materials will enrich the set of realistic
topological crystalline materials and stimulate related experiments. With
the demonstrated efficiency, our method \cite{TCP-Paper-1} can be employed
for a large-scale systematic search of the entire materials database, which
could lead to the discovery for many more new topological materials.
\begin{acknowledgements}
FT and XGW were supported by National Key R\&D Program of China (No.\ 2017YFA0303203 and 2018YFA0305700), the NSFC (No.\ 11525417, 51721001 and 11790311). FT was also supported by the program B for Outstanding PhD candidate of Nanjing University. XGW was partially supported by a QuantEmX award funded by the Gordon and Betty Moore Foundation's EPIQS Initiative through ICAM-I2CAM, Grant GBMF5305 and by the  Institute of Complex Adaptive Matter (ICAM).
AV is supported by NSF DMR-1411343, a Simons Investigator Grant, and by the ARO MURI on topological insulators, grant W911NF- 12-1-0961.
\end{acknowledgements}


\begin{thebibliography}{99}
\bibitem{review-1} M. Z. Hasan and C. L. Kane, Rev. Mod. Phys. \textbf{82},
3045 (2010).

\bibitem{review-2} X. L. Qi, and S. C. Zhang, Rev. Mod. Phys. \textbf{83},
1057 (2011).

\bibitem{rmp} C.-K. Chiu, J. C. Y. Teo, A. P. Schnyder and S. Ryu, Rev. Mod.
Phys. \textbf{88},035005 (2016).

\bibitem{TCI-1} L. Fu, Phys. Rev. Lett. \textbf{106}, 106802 (2011).

\bibitem{TCI-2} R.-J. Slager, A. Mesaros, V. Juri\u{c}i\'{c} and J. Zaanen,
Nature Phys. \textbf{9}, 98 (2013).

\bibitem{TCI-3} Y. Ando and L. Fu, Annu. Rev. Condens. Matter Phys. \textbf{6%
}, 361 (2015).

\bibitem{MCI} T. H. Hsieh, H. Lin, J. Liu, W. Duan, A. Bansil and L. Fu,
Nat. Commun. \textbf{3}, 982 (2012).

\bibitem{electric multipole insulators} W. A. Benalcazar, B. A. Bernevig, T.
L. Hughes, Science \textbf{357}, 61 (2017).

\bibitem{High-order-1} F. Schindler, A. M. Cook, M. G. Vergniory, Z. Wang,
S. S. P. Parkin, B. A. Bernevig and T. Neupert, Science Adv. \textbf{4},
eaat0346 (2018).

\bibitem{High-order-2} W. A. Benalcazar, B. A. Bernevig, and T. L. Hughes,
Phys. Rev. B. \textbf{96}, 245115 (2017).

\bibitem{High-order-3} Z. Song, Z. Fang, and C. Fang, Phys. Rev. Lett.
\textbf{119}, 246402 (2017).

\bibitem{High-order-4} J. Langbehn, Y. Peng, L. Trifunovic, F. von Oppen,
and P. W. Brouwer, Phys. Rev. Lett. \textbf{119}, 246401 (2017).

\bibitem{hourglass} Z. Wang, A. Alexandradinata R. J. Cava and B. A.
Bernevig, Nature \textbf{532}, 189 (2016).

\bibitem{nodal-chain} T. Bzdu\v{s}ek, Q. Wu, A. R\"{u}egg, M. Sigrist and A.
A. Soluyanov, Nature \textbf{538}, 75 (2016).

\bibitem{New-ferimon} B. Bradlyn, J. Cano, Z. Wang, M. G. Vergniory, C.
Felser, R. J. Cava, B. A. Bernevig, Sciences \textbf{353}, aaf5037 (2016).

\bibitem{wallpaper-Kane} J. Kruthoff, J. de Boer, J. van Wezel, C. L Kane
and R. J. Slager, Phys. Rev. X {\textbf{7}}, 041069 (2017).

\bibitem{Po} H. C. Po, A. Vishwanath, H. Watanabe, Nat. Commun. \textbf{8},
50 (2017).

\bibitem{QCT} B. Bradlyn, L. Elcoro, J. Cano, M. G. Vergniory, Z. Wang, C.
Felser, M. I. Aroyo and B. Andrei Bernevig, Nature \textbf{547}, 298 (2017).

\bibitem{Po-2} H. Watanabe, H. C. Po, and A. Vishwanath, arXiv:1707.01903
(2017).

\bibitem{wallpaper} B. J. Wieder, B. Bradlyn, Z. Wang, J. Cano, Y. Kim,
H.-S. D. Kim, A. M. Rappe, C. L. Kane and B. A. Bernevigm, arXiv:1705.01617
(2017).

\bibitem{Fu-Kane} L. Fu and C. L. Kane, Phys. Rev. B \textbf{76}, 045302
(2007).

\bibitem{FangFu} C. Fang and L. Fu, arXiv:1709.01929 (2017)

\bibitem{TCI-4} J. Langbehn, Y. Peng, L. Trifunovic, F. von Oppen, and P. W.
Brouwer, Phys. Rev. Lett. \textbf{119}, 246401 (2017).


\bibitem{TCI-6} W. A. Benalcazar, B. A. Bernevig, and T. L. Hughes, Phys.
Rev. B \textbf{96}, 245115 (2017).

\bibitem{Khalaf} E. Khalaf, H. C. Po, A. Vishwanath, and H. Watanabe,
arXiv:1711.11589 (2017).

\bibitem{Khalaf-1} E. Khalaf, arXiv:1801.10050 (2018).

\bibitem{Max} Max Geier, Luka Trifunovic, Max Hoskam and Piet W. Brouwer,
arXiv: 1801.10053 (2018).

\bibitem{Fang-1} C. Fang, M. J. Gilbert and B. A. Bernevig, Phys. Rev. B
\textbf{86}, 115112 (2012).

\bibitem{Fang-2} C. Fang and L. Fu, Phys. Rev. B \textbf{91}, 161105 (2015).

\bibitem{Fang-3} Z. Song, T. Zhang, Z. Fang and C. Fang, arXiv: 1711.11049
(2017).

\bibitem{Fang-4} C. Fang, Z. Song and T. Zhang, arXiv:1711.11050 (2017).

\bibitem{Weyl-1} X. Wan, A. M. Turner, A. Vishwanath, and S. Y. Savrasov,
Phys. Rev. B \textbf{83}, 205101 (2011).

\bibitem{Weyl-2} H. Weng, C. Fang, Z. Fang, B. A. Bernevig, and X. Dai,
Phys. Rev. X \textbf{5}, 011029 (2015).

\bibitem{Weyl-3} S.-M. Huang, S.-Y. Xu, I. Belopolski, C.-C. Lee, G. Chang,
B. Wang, N. Alidoust, G. Bian, M. Neupane, C. Zhang, S. Jia, A. Bansil, H.
Lin, and M. Z. Hasan, Nat. Commun. \textbf{6}, 7373 (2015).

\bibitem{Ca2As} X. Zhou, C.-H Hsu, T.-R. Chang, H.-J. Tien, Q. Ma, P.
Jarillo-Herrero, N. Gedik, A. Bansil, V. M. Pereira, S.-Y. Xu, H. Lin, and
L. Fu, arXiv:1805.05215 (2018).

\bibitem{ICSD} M. Hellenbrandt, Crystallography Reviews \textbf{10}, 17
(2004).

\bibitem{TCP-Paper-1} F. Tang, H. C. Po, A. Vishwanath, and X. Wan,
arXiv:1805.07314 (2018).


\bibitem{Pt3Ge} M. Ellner and B. Predel, Zeitschrift fuer Metallkunde
\textbf{51}, 327 (1960).

\bibitem{C} B. Vadlamani, K. An, M. Jagannathan and K. S. R. Chandran, J.
Electrochemical Society \textbf{161}, A1731 (2014).

\bibitem{Graphene-RMP} A. H. Castro Neto, F. Guinea, N. M. R. Peres, K. S.
Novoselov, and A. K. Geim, Rev. Mod. Phys. \textbf{81}, 109 (2009).

\bibitem{Graphene-gap} Y. Yao, F. Ye, X.-L. Qi, S.-C. Zhang, and Z. Fang,
Phys. Rev. B 75, 041401 (2007).

\bibitem{Kane-2005} C. L. Kane and E. J. Mele, Phys. Rev. Lett. \textbf{95},
226801 (2005).



\bibitem{graphite-gap} N García, P Esquinazi, J Barzola-Quiquia and S Dusari, New J. Phys. 14 053015 (2012).

\bibitem{Au4Ti} P. Pietrokowsky, Journal of the Institute of Metals \textbf{%
90}, 434 (1962).

\bibitem{Ti2Sn} C. Colinet, J. C. Tedenac and S. G. Fries, Comouter coupling
of phase diagrams and thermochemistry \textbf{33}, 250 (2009).

\bibitem{AgXF3} R. H. Odenthal and R. Hoppe, Monatshefte fuer Chemie \textbf{%
102}, 1340 (1971).

\bibitem{AgAsSr} A. Mewis, Zeitschrift fuer Naturforschung, Teil B.
Anorganische Chemie, Organische Chemie \textbf{33}, 983 (1978).

\bibitem{AgAsBa} A. Mewis, Zeitschrift fuer Naturforschung, Teil B.
Anorganische Chemie, Organische Chemie \textbf{34}, 1373 (1979).

\bibitem{MgB2} J. Nagamatsu, N. Nakagawa, T. Muranaka, Y. Zenitani, and J.
Akimitsu, Nature \textbf{410}, 63 (2001).
\end{thebibliography}
\end{document}